# THE LARGE-SCALE ELECTROCHEMISTRY AND POSSIBLE CONSEQUENCES OF *DEEP IMPACT* MISSION TO TEMPEL 1

E.M.Drobyshevski, E.A.Kumzerova, and A.A.Schmidt

A.F.Ioffe Physico-Technical Institute, Russian Academy of Science,
194021 St.Petersburg, Russia. E-mail: emdrob@mail.ioffe.ru

The assumption that short-period (SP) comets are fragments of massive icy envelopes of Ganymede-like bodies saturated by products of ice electrolysis that underwent global explosions provides a plausible explanation of all known manifestations of comets, including the jet character of outflows, the presence of ions in the vicinity of the nucleus, the bursts and splitting of cometary nuclei (which require energies of ~10-100 GJ and even more) etc., with solar radiation initiating burning of the products of electrolysis in the nucleus.

As shown persuasively by numerical simulation carried out in hydrodynamic approximation, the shock wave initiated by the *Deep Impact* impactor in the cometary ice saturated originally by the electrolysis products $2H_2 + O_2$ is capable of activating under certain conditions exothermal reactions (of the type $O_2 + H_2$ + organics → $H_2O$ + CO + HCN + other products of incomplete burning of organics including its light and heavy pyrolyzed compounds, soot ($C_2$, $C_3$), etc.), which will slow down shock wave damping (forced detonation) and increase many times the energy release. As a result, the size of the crater may exceed the volume determined by the kinetic energy of the impact while remaining nevertheless within the expected limits, which will lead to wrong conclusions concerning the properties of the nucleus. The key information will be contained in ejections and subsequent outflows from the crater. Their gaseous component should contain, if internal sources of energy are initiated, large amounts of products of high-temperature oxidation reactions occurring under shortage of oxygen. The total energy of ejections may be expected to exceed noticeably 10 GJ, for which purpose one should try to measure not only the velocity and temperature but their mass as well (but, in view of the uncertainties in estimates of the target density, not from the crater volume). The most spectacular result of a hit in the area of excess concentration of electrolysis products in the nucleus (which is most likely near the region of a jet outflow) would be breakup of the latter into fragments with a kinetic energy exceeding that of the impactor.

*Key Words:* collisional physics; comets, composition, origin; detonation; electrolysis; ices.

## 1. INTRODUCTION. *DEEP IMPACT* ACTIVE EXPERIMENT

The *Deep Impact* mission is scheduled to culminate on July 4, 2005 in an impact of a body 370 kg in mass and ~0.5 m in size on the nucleus of 9P/Tempel 1 comet with a velocity $u_{imp}$ = 10.2 km/s (kinetic energy $E_0$ = 19.4 GJ) (A'Hearn *et al.* 2005, and references therein; see also http://deepimpact.jpl.nasa.gov). The carrier spacecraft 500 km distant from the nucleus is expected to record the process itself and the consequences of the impact. Observations of the event from the Earth and by Hubble Space Telescope are also planned.

This active probing of the cometary nucleus is aimed at obtaining a basis for a sound judgment of its structure and composition. As a certain measure of our ignorance in this area may serve, for instance, estimates of the possible crater size, 50 < $d$ < 200 m (e.g., A'Hearn *et al.* 2005; Kruchynenko *et al.* 2005), which shows that its volume is predicted to within about one to two orders of magnitude. Melosh (1989, Ch. 7) also suggests this figure for the accuracy of crater volume prediction if no experimental data on impacts in the given conditions are available.

Clearly enough, not only the interpretation of the results obtained but even planning of the experiment itself, i.e., correct emphasis on certain aspects of the experiment and on treatment of the results obtained, depend in a large measure on the starting working hypotheses. In view of the fact that there is still more than one month left for possible adjustments and corrections, we are going to suggest here some predictions concerning possible consequences of the experiment, which are based on a self-consistent assumption that the vast majority of SP comets are actually products of global explosions of icy envelopes of Ganymede-like bodies saturated by products of the volumetric electrolysis of ice. Section 2 will expose briefly the main statements and achievements of this New Eruptive (Explosive) Cosmogony (NEC) of comets and other minor bodies of the Solar System. Section 3 will provide a short outline of present views concerning impact crater formation. Other conditions being equal, the crater volume is practically proportional to the energy released at the impact. If, as predicted by NEC, cometary ices do indeed contain, besides primitive organics and rock inclusions, products of the electrolysis of ice, more specifically, $O_2$ and $H_2$, dissolved in the form of clathrates, impact may initiate at the very least a non-self-sustained (forced) detonation (and thereafter, combustion), i.e., an additional release of energy. Calculations made in hydrodynamic approximation (Sec. 4) suggest that such an energy release may, under reasonable assumptions, exceed noticeably $E_0$, which would increase correspondingly the crater size. Section 5 suggests also some other consequences of the activation of an internal chemical source of energy, among them specific features in chemical composition and long duration of ejections and outflows of the dusty gas component from the crater, as well as the possibility for the nucleus to break up into fragments with kinetic energy $\geq E_0$.

## 2. ON THE ORIGIN OF COMETS

Discussions bearing on the goals and possible results of the *Deep Impact* mission are replete with statements that they should permit us to take a look into the past of the Solar System and formulate a judgment of the primordial material of which all its bodies had been made, including the comets themselves (e.g., Belton *et al*. 2005, and refs. therein). The latter are considered to be nothing more than building rubble, namely, agglomerates of rocks and ices of volatile compounds left over from the time of planetary accretion.

This traditional concept, however, comes in contradiction with practically all observational evidence. Consider here only a few, most obvious of them (Drobyshevski 1988b, 1997b, 1999, and refs. therein): (i) each comet has its own specific signature distinguishing it from others; (ii) ions and radicals were observed in the coma, in the immediate vicinity of the nucleus, much closer, in fact, than it would follow from the assumption of their photolytic formation from some hypothetical "parent" molecules (particularly remarkable in this respect is the detection of atomic and ionic carbon, C and $C^+$); (iii) the origin of CHON dust is unclear; (iv) the laws of conservation appear to be violated; indeed, an extremely small fraction of the surface area of the nucleus (about 5-10%) receiving an as small a part of incident solar energy releases large amounts of gases and dust in the form of jets, some of the jets being active on the night side of the nucleus as well; (v) in about 5% of cometary apparitions their nucleus splits into fragments flying away with velocities of up to 1-10 m/s (Sekanina 1982), and so on. This list could be continued further; suffice it to recall the anomalous long-lasting (for weeks!) breakups of P/Shoemaker-Levi 9 into fragments, different in composition, after an encounter with Jupiter with its enormously strong magnetic field, which can hardly be accounted for by tidal effects as this was supposed by some authors (e.g., Asphaug and Benz, 1994; Sekanina *et al*. 1994) (see discussion in Drobyshevski 1997b), and so on. No physically reasonable and non-contradictory answers have thus far being supplied to these questions. The only thing left to the proponents of the condensation-sublimation concept is an endless construction of ever new and frequently mutually excluding hypotheses based quite frequently on unobservable factors (amorphous ice, ejection of grains because of the (dirty) ice cracking under thermal impact (with this "impact" lasting years!), conservation in ice of energy-excessive chemical compounds and radicals created by cosmic rays or of bubbles of compressed gas, etc.).

The idea of ejection of comets from planetary bodies dates back to Lagrange and during the 20$^{th}$ century it was actively advocated by Vsekhsvyatski (1967) as an "eruptive concept" appealing to volcanic processes on the Galilean satellites mainly. The New Eruptive (Explosive) Cosmogony (NEC) of comets is based only on one well established electrochemical fact (Decroly *et al*. 1957; Petrenko and Whitworth 1999), namely, the possibility of electrolytic decomposition of ice in the solid phase (Drobyshevski 1980a; Drobyshevski *et al*. 1995). Volumetric electrolysis occurred in the icy envelopes with embedded carbonaceous and rock inclusions on moon-like bodies of the type of the outer Galilean satellites as they moved originally in the strong ancient magnetic field of planets or of the solar wind, which generated in them currents of up to ~$10^2$ MA. The electrolysis products $2H_2 + O_2$ subjected to pressures of up to $p \sim 0.1$-$1$ GPa build up in ice in the form of clathrates, a stable solid solution.

That an oxygen-based clathrate does exist at $T \approx 271$ K and pressures of ~10 MPa was demonstrated by van Cleef and Diepen in 1965. The oxygen content in a clathrate may amount to one $O_2$ molecule per ~6 $H_2O$ molecules. At $T \approx 120$ K, the oxygen clathrate is stable at $p \sim 20$ kPa (Byk *et al*. 1980). As for hydrogen, it was believed until quite recently that, similar to He and Ne, it cannot persist in clathrate structures at low pressures and should escape from the ice by diffusion. The situation changed only a year ago, when Mao and Mao (2004) reported existence of a high-pressure $H_2(H_2O)_2$ clathrate that holds 5.3 wt. % hydrogen at $T < 140$ K even at such a low pressure as $p \sim 100$ kPa (the last figures could also shed new light on some other features of cometary activity).

Saturation of ice with the products of electrolysis, $2H_2 + O_2$, up to concentrations of ~15 wt. % makes it capable of detonation. This is, however, a relatively weak, not a high explosive mixture. The detonation velocity in it is only $D \approx 5$ km/s, with pressure behind the detonation wave front $p_D \approx 5$ GPa (Drobyshevski 1986) (to compare with $D \approx 7$ km/s and $p_D \approx 30$ GPa for a standard TNT-type explosive (Baum *et al*. 1975)). Therefore, explosion of electrolyzed ice should not bring about crushing of unexploded fragments and loss of the gases dissolved in them.

Detonation can be initiated by a strong enough meteoroid impact. A global off-center explosion of the ice envelope of a moon-like body should shed off a substantial part (10-90%) of the ice (the actual fraction depends on the mass of the body) (Drobyshevski 1980b; Drobyshevski *et al*. 1994).

This approach permits one to explain and relate many astrophysical aspects, starting with the origin and properties of asteroids (Drobyshevski 1980a, 1997a; Drobyshevski *et al*. 1994) and of many small planetary satellites (Agafonova and Drobyshevski 1985; Drobyshevski 1988a), specific features in the structure and differences of the Galilean satellites (Drobyshevski 1980b), of Titan with its eccentricity and thick atmosphere and Saturn's rings (Drobyshevski 2000, and refs. therein), and ending with comets and the fine features of their manifestations and chemistry (Drobyshevski 1988b). A number of predictions made on the basis of this concept have been confirmed (a thing the traditional hypotheses cannot boast of), while others are still waiting for confirmation (Drobyshevski 2000, and refs. therein). NEC leads to certain conclusions concerning localization of conditions favorable for the origin of life (Drobyshevski 2002), while on the other hand substantiates the priority of exploration of comets for testing the NEC itself and, the last but not the least, argues convincingly for the need of sending missions to Callisto to test the possibility of explosion of its ices, which would provide a real threat to the very existence of Mankind (Drobyshevski 1999).

Viewed from the standpoint of NEC, comets are fragments of surface layers of the exploded icy envelopes. These fragments contain, besides primitive organics and rock inclusions, also $O_2$ and $H_2$, products of electrolysis dissolved in the ice (note that some critics of the NEC (e.g., Shulman 2000), in referring negligently to our publications, retort that $O_2$ and $H_2$ form in the cometary nuclei themselves in their interaction with interplanetary magnetic fields, or that we have in mind volcanic ejections from satellites resulting from $2H_2 + O_2$ explosions in volcanoes; being physically impossible, these processes are not mentioned in our papers at all). A fairly small input of additional energy (through insolation, meteoroid impact etc.) can initiate combustion and even explosions in such ices. One has to bear in mind that, being a product of geochemical differentiation and geological processes on the parent planets, these ices are, as a rule, non-uniform, have inclusions and a layered structure.

An impartial look reveals that all of the available observational evidence, including the above-mentioned facts, which are quite often referred to as "mysterious" and "incomprehensible", finds readily explanation within NEC without invoking any new hypotheses.

Recent data suggest, in particular, that a large part of the fairly well studied nuclei have an elongated shape (for instance, P/Halley, P/Borelli, P/Tempel 1, etc) (Jewitt *et al*. 2003), a feature characteristic of fragments originating from explosions of much larger bodies with geologically evolved structures (when their fragments of irregular shape are accelerated by a drag of the expanding gaseous products of the explosion more effectively and so escape the parent body easier), rather than of the products of accretion or, conversely, of collisions or collisional erosion. As for the quasi-spherical nucleus of P/Wild 2, the interesting features of its topography, including the irregular shape of the depression with abrupt walls, can be better understood if one takes into account the possibility of local burnout of isolated inclusions and layers enriched in combustible components.

## 3. IMPACT CRATER FORMATION.
## STRAIGHTFORWARD ESTIMATES FOR TEMPEL 1

Copious literature deals with the formation of craters by impact and explosions (e.g., Stanyukovich 1971; Roddy *et al*. 1976; Bazilevski *et al*. 1983; Anderson 1987; Melosh 1989). Straightforward considerations (see, e.g., Bazilevski *et al*. 1983; Melosh 1989) suggest that the crater volume $V$ should be proportional to the energy $E$ released in an explosion (or impact) and inversely proportional to the energy $q$ absorbed, on the average, by a unit volume of target material:

$$V = \chi E/q , \qquad (1)$$

where $\chi$ is a coefficient determined empirically ($\chi \sim 10^{-1} < 1$, which is due to one of dissipation mechanisms of many being, as a rule, considered dominant). Based on experimental data, one frequently assumes for the crater depth $h \approx d/4$, so that $V \approx \pi d^3/32$.

In the energy-based approach, one singles out usually two main modes determining the crater size, namely, the strength and gravitational ones.

We start with the last one. Here the order of magnitude for $q$ is defined as $q \approx \rho g d$ (where $\rho$ is the target material density, and $g$ is the acceleration of gravity), which yields

$$d \approx (32 \chi E / \pi g \rho)^{1/4}. \qquad (2)$$

For $E = 20$ GJ, $\rho = 1$ g cm$^{-3}$, $\chi = 0.1$, and $g = 10^{-4} g_0$ we obtain $d \approx 400$ m.

If the impact energy is absorbed primarily by plastic deformation of the target material, $q = Y$, i.e., the elastic limit of the material. For rough estimates one usually sets $Y =$ const; for granite $Y \approx 100$ MPa, for solid ice $Y \approx 17$ MPa at 257 K and $Y \approx 34$ MPa for $T = 81$ K (e.g., Lange and Ahrens 1987). Whence, assuming the nucleus of the comet to consist of solid ice at $T = 120$ K (i.e., $Y \approx 30$ MPa), we arrive at

$$d \approx (32 \chi E / \pi Y)^{1/3} \approx 10 \text{ m}, \qquad (3)$$

which is substantially less than the estimate obtained in the gravitational approach.

One can hardly assume, however, that at high strain rates matter would behave as a continuous medium. Only a small fraction of its volume bears the main load, and this reduces strongly $q$ (this may partially account for the coefficient $\chi$ found experimentally being <1). Indeed, in the presence of high shear strain rate gradients adiabatic shear bands appear, which initiates formation of a gas (or even plasma) phase even in the conditions where volume-averaged consideration would suggest the very onset of liquid phase formation. This relates to such high-plasticity materials as metals. Therefore, ejection out of craters of solid blocks (by the gas component) occurs with a higher velocity and efficiency (Drobyshevski 1995). Another mechanism reducing the effective value of $q$ is the brittle fracture of material by the shock wave that crosses it (Stanyukovich 1971). Here the volume energy expended to crush monolithic material into large blocks is likewise much lower than that needed to shift molecular layers with respect to one another. This is particularly typical of brittle material with a low elastic deformation threshold (ceramics, rocks etc.). As follows from calculations of Nolan *et al.* (1996), in large scale impacts (impactor size >5 m for $u_{\text{imp}} > 5$ km/s) $q$ drops to such low levels that in real conditions crater formation on bodies already as small as ≥1-10 km occurs in the gravitation regime, with ejection of large fragments at low velocities determined by elastic stress relaxation.

There are more sophisticated approaches to estimation of the consequences of impacts (with inclusion of momentum transfer, with the use of the so-called $\pi$ parameters, etc.). All of them, however, are based on empirical normalizations, which are determined each time for impacts of a given class, and are capable of estimating the crater volume, other conditions being equal, at best to within an order of magnitude (Melosh 1989).

Even the rough estimates presented above demonstrate that the expected size of the crater on Tempel 1 is determined by such a large set of unknown parameters and lies, therefore, within such a large range, that even an accurate enough measurement of the crater diameter and depth will hardly permit a reliable judgment of the material and structure of the cometary nucleus or shed light on its composition (for ice - monolithic or porous, amorphous, low or high pressure phase; fraction, state of dispersion, stratification etc. of rock and organic inclusions; presence and structure of nonvolatile crust and so on). We might add to this list one more parameter, namely, internal source of energy. Our main hope is placed therefore on an analysis of the composition of gas jet components, their duration, more specifically, on the formation of long-lived gas/dust jets out of the crater or of its impact-perturbed surroundings, measurement of the total kinetic energy of the outflows and ejections and, possibly, of large fragments of the nucleus.

## 4. IMPACT CONSEQUENCES WITH INCLUSION OF FORCED DETONATION OF ELECTROLYZED ICES

As already mentioned, all mysterious and bizarre manifestations of comets are readily accountable for by assuming that their ices are saturated (non-uniformly) by the products of electrolysis up to a concentration $α ≈ 15$-$20$ wt. %. Our early estimates (Drobyshevski 1986) suggest that such a uniform solid solution is capable at $α ≈ 17$ wt. % of stable detonation at the initial temperature $T_0 ≥ 145$ K, i.e., at a temperature reached in icy envelopes of Galilean satellites at a depth of tens of km. We did not consider reaction kinetics behind the shock front. As a criterion of detonation, i.e., of instantaneous liberation of energy initiated by shock compression of material in the shock wave we accepted the temperature $T_c = 900$ K reached in the products of detonation (for conventional explosives, the lowest temperature $T_c = 700$-$900$ K, see pp. 165-166 and 195-197 in Baum *et al.* 1975). Therefore, at $T_0 ≈ 120$ K (distance to Jupiter) stable detonation throughout the volume of the body is impossible, as soon as even under the assumption of the reactions being fully completed the temperature would be lower than $T_c = 900$ K.

At the instant a body hits the ice with $u_{imp} \sim 10$ km/s the temperature reaches $\sim 10^4$ K. The energy of *Deep Impact* (~20 GJ) is only enough to melt less than 38 t of conventional ice which was originally at $T_0 = 120$ K. In actual fact, however, this figure will be smaller (see below), because part of the energy in the area of direct contact will be expended to heat material to the plasma state and will be driven by the decaying shock wave into the bulk of the target while only partially transforming into the kinetic energy of directed motion of material. If, however, as a result of the initial impact the temperature behind the shock front $T ≥ T_c$, the mixture $2H_2 + O_2$ + organics will be able to react with liberation of energy, thus sustaining the shock wave and slowing down its damping. We deal here with the phenomenon of forced detonation (Baum *et al.* 1975) where liberation of energy behind the shock front is insufficient to make it stationary. In the calculations that follow we have accepted persistence of $2H_2 + O_2$ stoichiometry, bearing in mind the existence at low temperatures and pressures of stable hydrogen and oxygen clathrate hydrates discussed in Sec.2

Our purpose will in this case be estimation of the additional energy that will be added to 20 GJ of the impact, as well as of the mass which will be involved in forced detonation. The latter figure is important, because it determines the amount of the products of high-temperature exothermal reaction $O_2 + H_2$ + organics which hopefully will be estimated in observations of the consequences of the impact. The analysis below was conducted along the lines of the ideology formulated by Drobyshevski (1986).

Interaction of an impactor with the plane surface of the cometary nucleus was numerically simulated in terms of one-dimensional, spherically symmetrical, formulation. We considered evolution of a spherical layer in half-space, with an impactor-initiated pressure pulse specified at a point ($x = 0$) on the surface of this layer.

Propagation of a shock wave was described by standard equations of the mass, momentum, and energy conservation in integral form (Godunov 1976):

$$\oint_\Gamma a\,dx - b\,dt = \iint_\Omega \frac{2}{x}(f-b)\,dx\,dt, \qquad (4)$$

$$a = \begin{bmatrix} \rho \\ \rho u \\ \rho(e+u^2/2) \end{bmatrix}, \quad b = \begin{bmatrix} \rho u \\ p+\rho u^2 \\ \rho(e+u^2/2)u+pu \end{bmatrix}, \quad f = \begin{bmatrix} 0 \\ p \\ 0 \end{bmatrix},$$

where the internal energy of the medium $e$ includes the possible ice-water-vapor phase transitions, as well as the energy released in exothermal reactions behind the shock front.

At this time, one could hardly venture a guess on thermodynamic properties of the solid solution (clathrate?) of $2H_2 + O_2$ in dirty cometary ice, all the more that it contains ~10% of primitive organics. Therefore, to describe the properties of matter behind the shock front and close the system (4), we assumed an equation of state for water in the form (Baum *et al.* 1975; Drobyshevski 1986):

$$p(\rho,T) = \frac{2.992\cdot 10^8 (r^{7.3}-1)}{1+0.7(r-1)^4}(1-0.012 R^2 f) + 4.611\cdot 10^5 Rf(T-273), \qquad (5)$$

$$f = \frac{1+3.5r-2r^2+7.27r^6}{1+1.09r^6}, \quad r = \frac{\rho}{\rho_*}, \quad \rho_* = 1000 \text{ kg m}^{-3}.$$

The internal energy of water can be written as

$$e_w(\rho,T) = 6.3\cdot 10^6 \left(1-\frac{1}{r}\right)\left(0.71-\frac{1}{r}\right)r^{4/3}\left[1-2r\exp(-r^2)\right] + 3651.28\,T + \text{const.} \qquad (6)$$

Internal energy of ice was represented in the form

$$e_i = c_v T_i \text{ J}, \qquad (7)$$

where the specific heat of ice depends on its temperature as

$$c_v = 7.7\,T_i \text{ J kg}^{-1}\text{ K}^{-1}. \qquad (8)$$

It was assumed that for pressure $p \geq 2\times 10^8$ Pa the ice-water phase transition temperature can be approximated by the relation

$$T_{iL} = 0.508\cdot 10^{-7} p + 243 \text{ K}, \qquad (9)$$

and the heat of the ice-water phase transition

$$L = 3\cdot 10^5 \text{ J kg}^{-1}. \qquad (10)$$

Oxidation reactions liberate energy at temperatures above a critical level $T_c$, which lies in the 700-900 K interval (Baum *et al*. 1975) and depends on $\alpha$, the mass content of the products of electrolysis, $2H_2 + O_2$, in the mixture

$$Q = 13.27\alpha \text{ MJ kg}^{-1}. \tag{11}$$

Because the binding energy of the $H_2$ and $O_2$ molecules with $H_2O$ molecules and with one another in ice is small compared to that between water molecules, it was assumed that the $2H_2 + O_2$ components with a mass fraction $\alpha$ are not involved in the energy-consuming phase transitions of water and behave as an ideal gas.

Equations (4) were solved for the region behind the shock wave. It was assumed to propagate with the velocity

$$D = \frac{1}{\rho_0} \left[ (p - p_0) / \left( \frac{1}{\rho_0} - \frac{1}{\rho} \right) \right]^{1/2}, \tag{12}$$

where $p_0$ and $\rho_0$ are the initial ice pressure and density (in Drobyshevski 1986, the square brackets were omitted by misprint), and the fluxes at the boundaries of the considered volume were determined by means of the Rankin-Hugoniot relation at the pressure jump, with inclusion of the possibility of a phase transition and an additional energy release:

$$e - e_0 = \frac{p + p_0}{2} \left( \frac{1}{\rho_0} - \frac{1}{\rho} \right). \tag{13}$$

In accordance with NEC, it was assumed, as this was done by Drobyshevski (1986), that the density of cometary ices is that of the high-pressure phases ($\rho_0 = 1183$ kg m$^{-3}$), which persist at low temperatures ($T_0 \sim 100$ K) and after removal of the load, say, at $p_0 = 10^5$ Pa. Decreasing $\rho_0$ to 920-1000 kg m$^{-3}$ affects only weakly the result (the available estimates of the low density of cometary nuclei, $\rho \sim 300$-500 kg m$^{-3}$, were obtained within pure sublimation models; hopefully, the *Deep Impact* observations of the ejected fragments will permit to evaluate the real mass and density of the nucleus (Belton *et al*. 2005, and refs. therein)).

The amplitude of the initial pressure pulse $p_{imp}$ can be roughly estimated from the following relation (Baum *et al*. 1975)

$$p_{imp} = \frac{\rho_{ice} c_{ice} \rho_{cop} c_{cop} u_{imp}}{\rho_{ice} c_{ice} + \rho_{cop} c_{cop}}, \tag{14}$$

where $\rho_{cop}$ and $\rho_{ice}$ are, accordingly, the densities of the impactor (copper) and cometary envelope (ice), $c_{cop}$ and $c_{ice}$ are the sonic velocities in these media, and $u_{imp}$ is the impact velocity. Whence one obtains for the initial pressure in the shock wave $p_{imp} \approx 40$ GPa.

The pulse duration was estimated as

$$p_{imp} S_{imp} t = m_{imp} u_{imp}, \tag{15}$$

where $m_{imp}u_{imp}$ is the impactor momentum and $S_{imp} \approx 0.3$ m$^2$ is the impactor area, which yields for the pulse duration

$$t = \frac{m_{imp} u_{imp}}{p_{imp} S_{imp}} \simeq 10^{-4} \text{ s}. \tag{16}$$

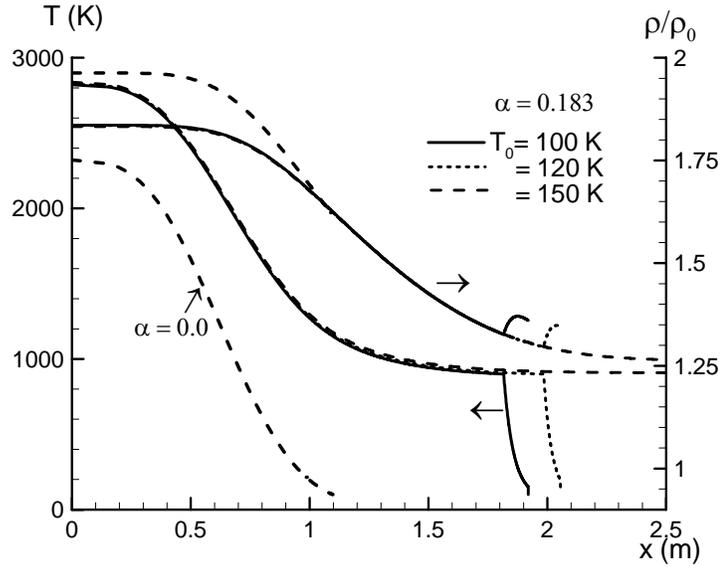

**FIG. 1a.** Temperature $T$ and density $\rho/\rho_0$ behind the shock wave calculated for $\alpha = 0$ and $\alpha = 0.183$ and the original ice temperature $T_0 = 100$, 120, and 150 K and plotted vs. distance $x$ from the point of impact. $T_c = 900$ K.

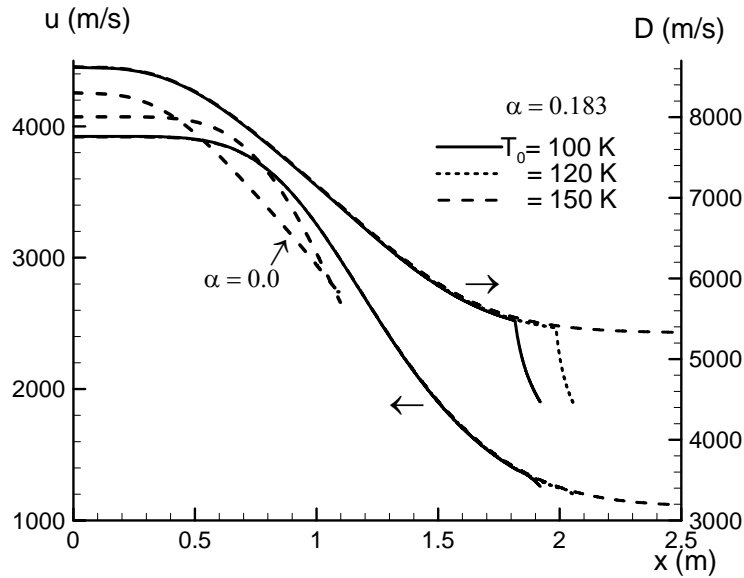

**FIG. 1b.** Shock wave velocity $D$ and velocity $u$ behind the shock wave calculated for $\alpha = 0$ and $\alpha = 0.183$ and the original ice temperature $T_0 = 100$, 120, and 150 K and plotted vs. distance $x$ from the point of impact. $T_c = 900$ K.

Calculation of the processes evolving with time behind the shock wave propagating away from the impact zone into the nucleus of P/Tempel 1 for $\alpha = 0$, as well as for $\alpha = 0.183$, a level high enough to sustain stable detonation ($T_c = 900$ K) at $T_0 = 150$ K, is illustrated in Figs. 1a,b.

We readily see that starting from $x = 1.81$ m from the point of impact for an initial ice temperature $T_0 = 100$ K, and from $x = 1.98$ m for $T_0 = 120$ K, the graphs plotting the variation of the parameters of material ($T, \rho/\rho_0, D, u$) behind the shock front undergo a break, because the shock wave begins to decay rapidly as a result of the excess energy no longer being released. Interestingly, because of a temperature drop due to the energy release ceasing, the density behind the shock wave even increases somewhat initially.

Figure 2 provides an idea of the total amount of additional energy liberated in the cometary material as a result of initiation in it of forced detonation. Significantly, this energy, under reasonable assumptions, may exceed noticeably the planned energy of *Deep Impact*.

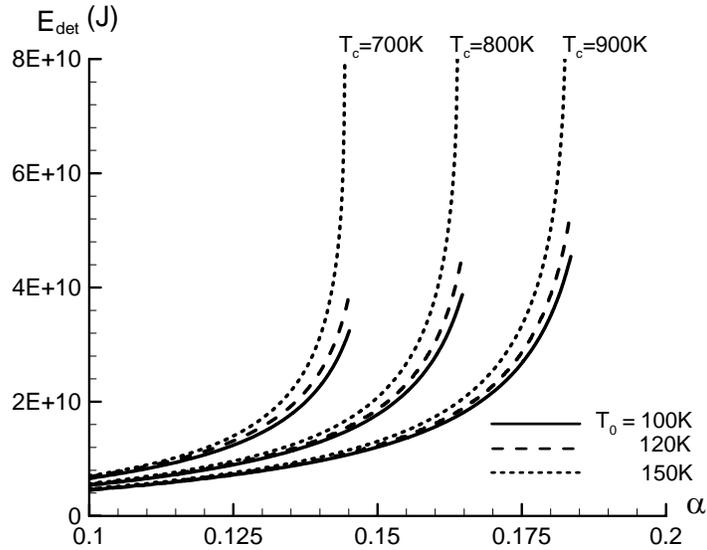

**FIG. 2.** Energy released by the shock wave plotted vs. $\alpha$, the $2H_2 + O_2$ content in ice, at various critical detonation temperatures $T_c$ and $T_0 = 100, 120,$ and $150$ K.

5. DISCUSSION OF POSSIBLE OBSERVATIONAL CONSEQUENCES OF THE IMPACT

As follows from the calculations (Fig. 2), the additional energy liberated in the electrolyzed ices of the comet as a result of their impact-initiated forced detonation exceeds the energy of the impact itself only about two- to threefold. This is too small a value to drive the crater size beyond the broad range of uncertainty of the predicted figures.

The impact itself with $E_0 = 20$ GJ could convert to water <38 t of pure ice (at $T_0 = 120$ K), evaporate <6.2 t, or heat <5.5 t of ice to vapor state at $T = 900$ K. In actual fact, taking into account the strongly non-uniform distribution of energy, part of which would be expended to overheat the material in the immediate vicinity of the impactor and the impactor itself, these figures should be scaled down by about a factor five (indeed, as follows from Fig. 1a, for $\alpha = 0$ only 0.925 t of material would be heated to $T \geq T_c = 900$ K). By contrast, in the case of detonation accompanied by liberation of chemical energy $E = 2.5E_0 = 50$ GJ, when release of additional energy will give rise to a kind of quasi-thermostatting behind the shock wave, the amount of ice, together with the organics it contains, vaporized and inhomogeneously heated to $T \geq T_c = 900$ K will be 19.2 t. This exceeds by more than an order of magnitude (~20 times) the values associated with the impact alone. Therefore, the total (thermal and kinetic) energy contained in the outflow of the gas and of the finely dispersed inclusions and products of some gas components' condensation in expansion in vacuum may turn out comparable to $E_0$.

Accordingly, the amount of the products of pyrolysis of the organics contained originally in ices should increase by an order of magnitude too.

What should increase by several orders of magnitude, is the amount of the products of combustion of these organics under deficiency of the oxydizer, which is determined by the relative amounts of these organics and of the clathrate hydrogen retained in the ices. These are, besides $H_2O$, the various "quasi-cometary" molecules such as $H_2$, $N_2$, CO, $CO_2$, HCN, $H_2S$, $H_2CO$, $CH_3OH$, light pyrolyzed hydrocarbons etc., as well as, on the one hand, their radicals and ions (however, in relatively low concentrations due to the shock-wave-caused reactions proceed at high pressure), and on the other, carbon-containing soot and CHON particles accounting for the "smoke" produced in incomplete combustion. It is essential that these components should form immediately at the instant of ejection rather than at a large distance from the nucleus, which could be assigned to subsequent photolysis. The existence of a well fixed point of reference, namely, the time of impact, distinguishes favorably this experiment from flyby observations of jets evolving continually from narrow discrete sources on the surface of the nucleus and carrying the low-pressure combustion products containing a great deal of radicals and ions. It would be instructive to monitor the practically instantly ejected cloud of the high pressure combustion products to see how photolytic and solar-wind-related processes will convert them into a substance of cometary coma and tail which is rich in radicals and ions.

One cannot exclude the possibility that forced detonation will transform to deflagration, i.e., detonation-initiated non-shock low-pressure combustion of material. In this case, the outflow of combustion products from the crater would tail out, and we will become witnesses to (i) formation of jets of material emanating even on the night side of the nucleus, and (ii) a gradual increase in crater size (not so fast as the one caused by the impact) accompanied by the appearance of burnt-out grooves where the content of combustibles was originally enhanced (recall the Stickney crater on the burnt-out Phobos (Drobyshevski 1988a) and possibly similar large craters on minor low-density bodies (Thomas 1999)).

Our analysis was necessarily restricted to an idealized model of impact on uniform ice. But, first, as already mentioned, cometary ices are products of geological processes in the parent Ganymede-like bodies. Therefore, their structure is spatially non-uniform; indeed, measuring many km in size, they have a complex structure and contain inclusions which were originally enriched or depleted in some minerals, including organics and products of electrolysis. There could even be meter-sized rock inclusions (Drobyshevski 1980a). Second, subsequent evolution of cometary nuclei including loss of volatiles gives rise to development of a non-uniform surface crust consisting of "sand" strengthened by pyrolyzed and cosmic ray processed organics.

It thus appears hardly possible to predict unambiguously the consequences of the impact. Impact on a boulder would bring about results radically different from those of an impact on a vein with an enhanced concentration of the products of electrolysis and organics, somewhere in the vicinity of a jet source. In the latter case, which would be of most interest for a scientist, one could conceive of a situation where waveguide properties of such a vein would drive detonation very deep into the nucleus and even culminate in its breakup into large fragments flying away from one another with $E \sim E_0$ or even greater, a case observed on many occasions.

ACKNOWLEDGMENTS

The authors are grateful to M.J.S.Belton for providing with preprints on the Deep Impact Project, to K.I.Churyumov for fruitful discussions, and to R.Horgan for drawing our attention to works by Mao and Mao on hydrogen clathrates.